\documentclass[lettersize,journal]{IEEEtran}
\usepackage{amsmath,amsfonts}
\usepackage{algorithmic}
\usepackage{algorithm}
\usepackage{array}
\usepackage[caption=false,font=normalsize,labelfont=sf,textfont=sf]{subfig}
\usepackage{textcomp}
\usepackage{stfloats}
\usepackage{url}
\usepackage{verbatim}
\usepackage{graphicx}
\usepackage{cite}
\usepackage{amsthm}
\usepackage{cases}
\usepackage{bm}
\usepackage{xcolor}
\newtheorem{remark}{Remark}
\newcommand{\rev}[1]{{\leavevmode\color{black}#1}}
\newcommand{\revv}[1]{{\leavevmode\color{black}#1}}
\newcommand{\fin}[1]{{\leavevmode\color{black}#1}}

\begin{document}

\title{A Foundation for Wireless Channel Prediction and Full Ray Makeup Estimation Using an~Unmanned Vehicle}

\author{{Chitra R. Karanam, and Yasamin~Mostofi,~\IEEEmembership{Fellow,~IEEE}}
\thanks{This work was supported in part by NSF RI award 2008449.}
\vspace{-0.3in}}%



\maketitle

\begin{abstract}
In this paper, we consider the problem of wireless channel prediction, where we are interested in predicting the channel quality at unvisited locations in an area of interest, based on a small number of prior received power measurements collected by an unmanned vehicle in the area. We propose a new framework for channel prediction that can not only predict the detailed variations of the received power, but can also predict the detailed makeup of the wireless rays (i.e., amplitude, angle-of-arrival, and phase of all the incoming paths).  More specifically, we show how an enclosure-based robotic route design ensures that the received power measurements at the prior measurement locations can be utilized to fully predict detailed ray parameters at unvisited locations. We then show how to first estimate the detailed ray parameters at the prior measurement route and then fully extend them to predict the detailed ray makeup at unvisited locations in the workspace. We experimentally validate our proposed framework through extensive real-world experiments in three different areas, and show that our approach can accurately predict the received channel power and the detailed makeup of the rays at unvisited locations in an area, considerably outperforming the state-of-the-art in wireless channel prediction.
\end{abstract}


\vspace{-0.1in}
\section{Introduction}\label{sec_intro}
Wireless channel prediction is a problem of considerable interest to the research community. Broadly speaking, \textit{channel prediction} is the prediction of the wireless channel signal strength (or the wireless channel quality) at various unvisited locations in an area of interest, based on a small number of prior channel power measurements in the area. Channel prediction is an integral part of many applications, including wireless router placement \cite{wei2017facilitating, wang2007efficient}, robotic path planning \cite{TRO19_MuralidharanMostofi, Autonomous19_MuralidharanMostofi}, and fingerprinting-based localization approaches \cite{sun2018augmentation}. These applications rely on channel prediction in order to predict and learn the spatially-varying wireless signal strength at unvisited locations, instead of going through the cumbersome process of collecting signal strength measurements at every location in the area of interest. 

In robotics, for instance, unmanned vehicles need to maintain connectivity among themselves, and/or to remote operators, in order to ensure a satisfactory task completion. The wireless channel signal strength, however, varies spatially, which can present a challenge for mobile robots to stay connected and provide the needed quality of service as they traverse the environment. Thus, in order to properly plan its actions and maintain the needed level of connectivity, it is crucial for an unmanned vehicle to predict the wireless channel at
unvisited locations in the workspace. Predicting the channel at unvisited locations is also important for non-robotic applications. For instance, consider the placement of a wireless router in an environment. If for a given router transmitter location, we can predict the corresponding channel over the area of interest, we can then properly optimize the placement of the router in order to achieve the desired signal strength map, without exhaustively measuring the channel over the whole space.  Similarly, fingerprinting-based localization applications can also benefit tremendously from channel prediction \fin{based on} only a sparse set of prior samples in the environment.

Given the importance of and the need for such a prediction framework, a number of different approaches have been proposed to solve the problem of wireless channel prediction in recent years.  In one category of work, objects are localized and ray tracing approaches are used to predict the channel. \cite{ho1993wireless}, for instance, utilizes an image-based ray tracing method along with prior information on building material and antenna patterns to predict the channel. The authors in \cite{wei2017facilitating, zhou2020robotic} use prior mmwave signals in order to first localize objects and then perform ray tracing. However, accurate detailed object localization/geometry estimation, using only RF signals, is itself a challenging problem and the subject of extensive research in the RF community \cite{karanam20173d, gonzalez2013cooperative, huang2014feasibility}.  Furthermore, these work have to assume that each object reflects uniformly in all the directions, which is not necessarily the case, or they need to exhaustively learn the reflection pattern of each object.  Finally, they largely assume primary (single-bounce) reflections.

In our past work \cite{malmirchegini2012spatial}, we have developed a Gaussian Processes (GP)-based approach for channel prediction, which has also been utilized and extended by others \cite{fink2013robust, caccamo2017rcamp, kalogerias2018spatially, ghaffarkhah2014dynamic, muralidharan2021communication}. This method models the underlying path loss and shadowing components with a Gaussian process and utilizes the corresponding spatial correlation to predict the channel \fin{at unvisited locations}.  While this approach is shown to provide a decent prediction performance and is used in a number of robotic applications, it cannot predict the multipath variations and as such will miss the detailed spatial changes of the channel (Sec.~\ref{sec_exp_results} will show the performance of this approach). Finally, machine learning has been utilized mainly for aerial channel prediction \cite{aldossari2019predicting, sotiroudis2021fusing, han2020power, krijestorac2020spatial, teganya2020data}. However, either only the path loss component is predicted and/or the trained network is only tested in a simulation environment. In general, training a Deep Neural Network requires a vast amount of data related to many different scenarios and training on small datasets results in limited generalizability. \fin{Overall, the existing work on channel prediction have also mainly focused on predicting the received signal power and not the detailed makeup of the rays.}

\fin{In this paper}, we are interested in providing a core understanding \fin{on} the mathematical foundation of channel prediction. More specifically, we propose a new framework for channel prediction that can predict the detailed makeup of the rays at unvisited locations in the area of interest, \fin{based on only a small number of prior received power measurements in the area.} \textbf{By detailed ray makeup, we mean the amplitude, angle of arrival, and phase of all the wireless rays that arrive at any given unvisited location in the workspace.}  As such, our approach not only predicts the detailed spatial variations of the channel but also reveals the detailed ray makeup at unvisited locations, providing additional important information for communication planning and optimization. We next discuss the main contributions of the paper in detail:

\begin{itemize}
    \item We propose a new framework for robotic channel prediction, where we utilize the mobility of a robot to collect a small number of prior wireless signal power measurements in the area of interest. We show how to design the route of the robot in order to optimize the locations/configuration of the prior measurement collection phase and capture the most important information that constitutes the makeup of the rays passing through the area. More specifically, we show how an enclosure-based route design can ensure that the ray information extracted at the prior locations can be utilized to predict detailed ray parameters at unvisited locations in the area. 
    \item We show how to estimate the full makeup of the signal rays that propagate through the area of interest. \fin{More specifically}, we show how we can utilize the methodically-designed prior measurement routes to fully estimate the detailed parameters that constitute any signal ray that arrives at the prior measurement routes. We then show how we can extend these extracted parameters and subsequently predict the full makeup of the rays (as well as the corresponding received signal power) at unvisited locations in the workspace. 
    \item We validate our proposed approach through extensive experiments in three different areas. Our results show that our approach can predict the key parameters of the signal rays and reveal their detailed makeup at unvisited locations. Furthermore, our results confirm that our approach can predict the detailed small-scale multipath variations that arise from reflection, diffraction or any other propagation phenomena in the area. We also compare the performance of our approach with the state-of-the-art, which shows how our approach can predict the detailed channel variations, which are \revv{not captured} by the state-of-the-art.  
\end{itemize}

The rest of the paper is organized as follows. In Sec.~\ref{sec_problem_formulation} we introduce the wireless signal propagation model and discuss the key parameters that make up the rays propagating through an area. In Sec.~\ref{sec_proposed_framework}, we discuss our proposed approach for analyzing the prior wireless measurements and subsequently predicting the full ray makeup at any unvisited location in the area of interest. In Sec.~\ref{sec_exp_results}, we extensively validate our proposed approach through experiments in three different areas on our campus and further compare with the state-of-the-art. We conclude in Sec.~\ref{sec_conclusions}. 

\vspace{-0.1in}
\section{Problem Formulation}\label{sec_problem_formulation}
\begin{figure}
    \centering
    \includegraphics[width=1\linewidth]{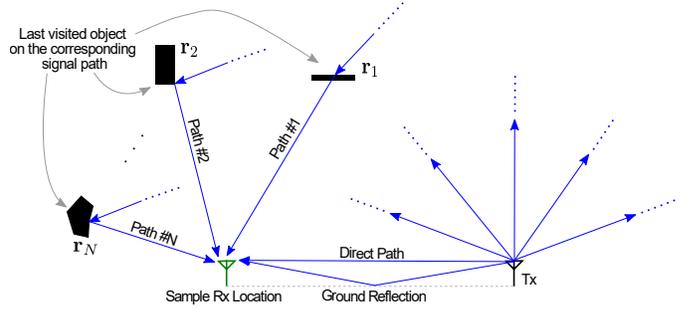}
    \caption{An example of our considered scenario with a fixed transmitter (Tx) that illuminates the area. The signal from the Tx propagates through the area, \revv{interacts with the objects}, and finally arrives at the receiver (Rx) antenna.}
    \vspace{-0.2in}
    \label{fig_general_scenario}
\end{figure}

Consider a scenario with a transmitter (Tx) and several objects located in an area, as shown in Fig.~\ref{fig_general_scenario}. The signal rays from the Tx interact with the objects in the area and undergo various propagation phenomena including reflection, absorption, and diffraction. These rays subsequently pass through the area, interacting constructively or destructively with each other to result in a spatially-varying channel with a varying signal power at different locations in the area. 

\begin{figure*}
\begin{minipage}{0.25\textwidth}
    \vspace{0.22in}
    \includegraphics[width=0.95\linewidth]{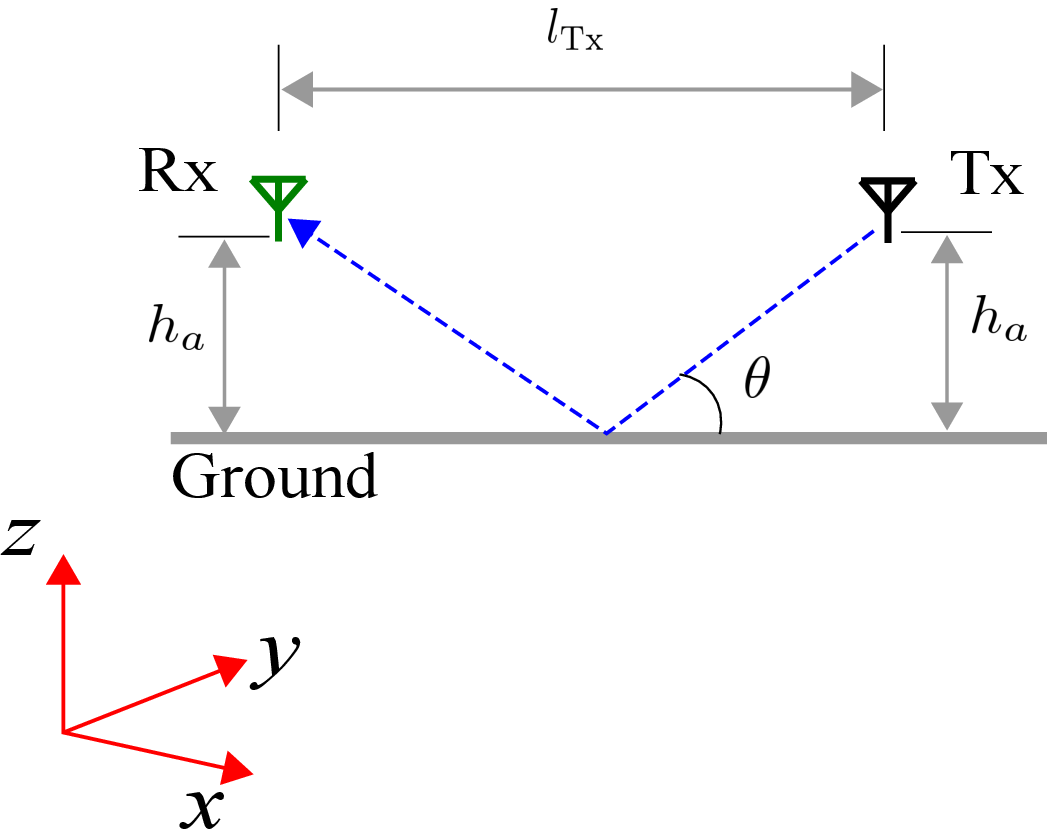}
    \caption{Signal path propagation from Tx to the ground to the Rx antenna.}
    \label{fig_ground_params}
    \end{minipage}\hfill
\begin{minipage}{0.28\textwidth}
    \centering
    \vspace{0.2in}
    \includegraphics[width=0.95\linewidth]{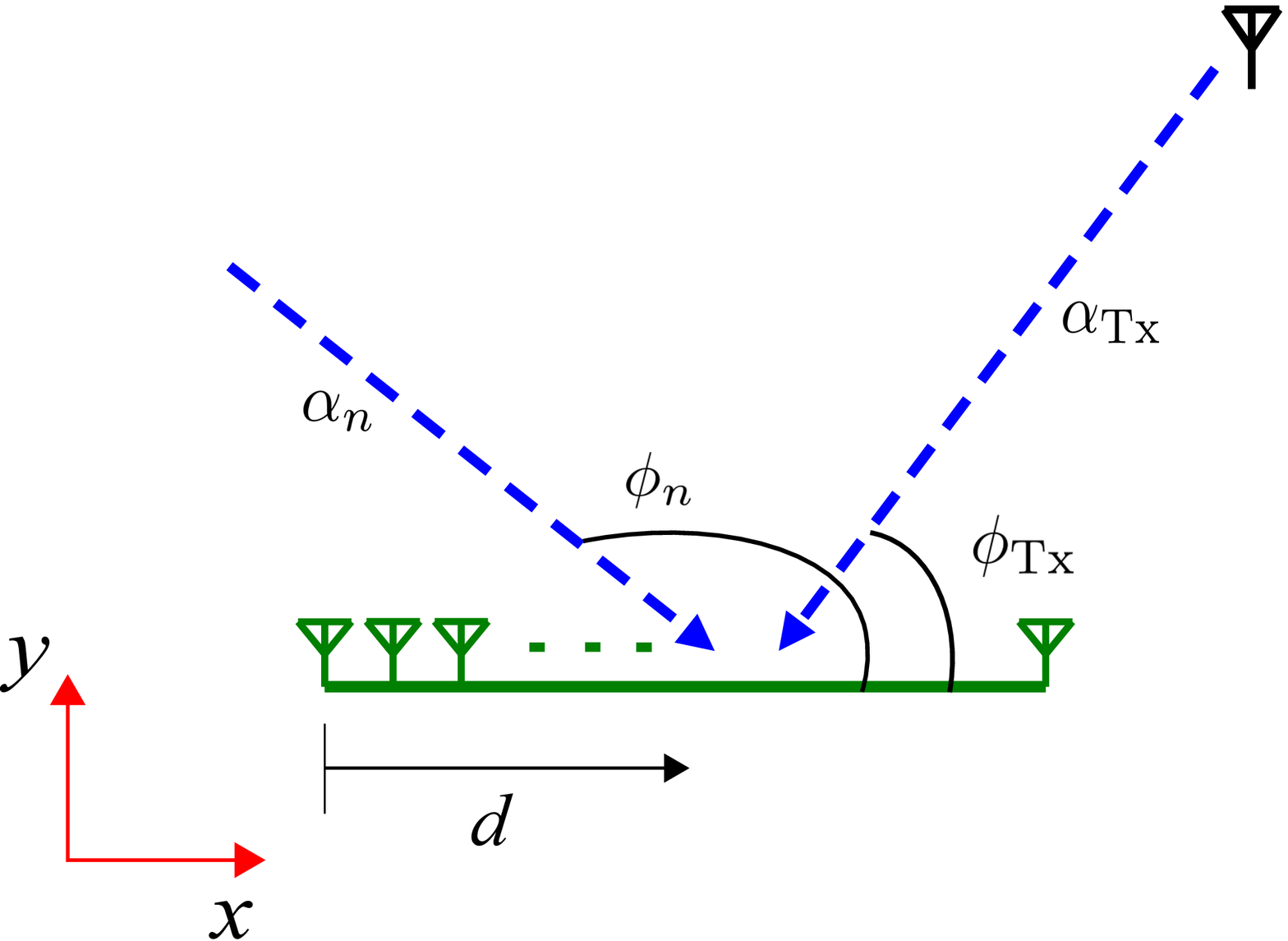}
    \vspace{-0.05in}
    \caption{An example top-view of the signal paths arriving at a receiver antenna array.}
    \label{fig_antenna_array}
\end{minipage}\hfill
\begin{minipage}{0.4\textwidth}
    \centering
    \includegraphics[width=0.85\linewidth]{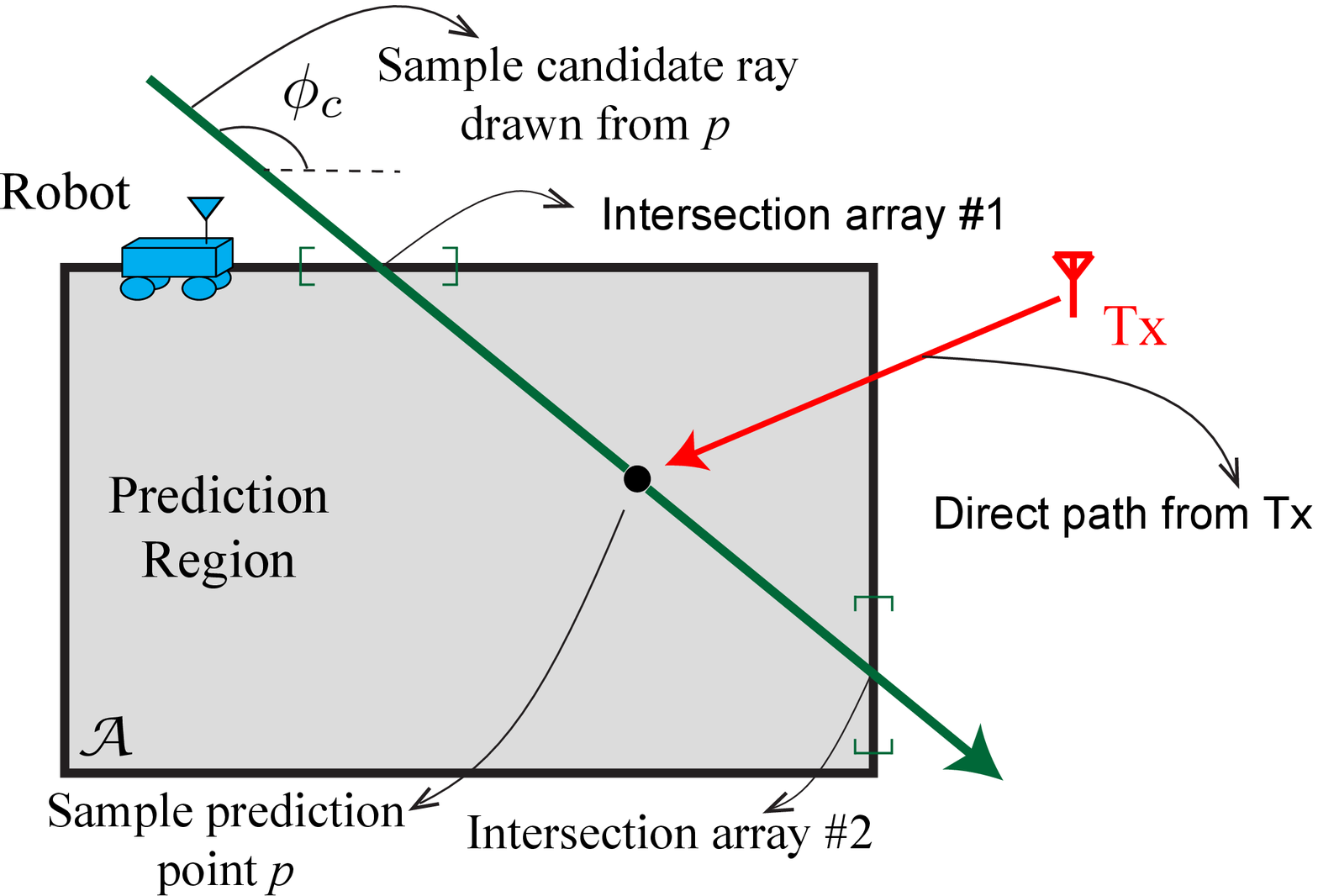}
    \vspace{-0.1in}
    \caption{Channel prediction scenario -- It is of interest to predict the detailed channel makeup at all the locations in region $\mathcal{A}$.}
    \label{fig_prediction_scenario}
\end{minipage}
\vspace{-0.15in}
\end{figure*}

Consider a sample receiver (Rx) point, as shown in Fig.~\ref{fig_general_scenario}. Suppose that the Tx and Rx are located at $\mathbf{r}_\text{Tx}$ and $\mathbf{r}_{\text{Rx}}$ respectively. The complex baseband received signal at any such receiver location in the area can then be written as,
\vspace{-0.05in}
\begin{equation}\label{eq_point_signal}
c(\mathbf{r}_{\text{Rx}}) = \alpha_{\text{Tx}} e^{j \frac{2\pi}{\lambda} l_{\text{Tx}}} + \alpha_{g} e^{j\frac{2\pi}{\lambda} l_{g}} + \sum_{n=1}^N \alpha_{n} e^{j \frac{2\pi}{\lambda} l_{n}} + \eta(\mathbf{r}_{\text{Rx}}),
\end{equation}
where $\alpha_{\text{Tx}} = \frac{\lambda P_t G_t G_r}{4\pi l_{\text{Tx}}}$ is the amplitude of the direct path from the Tx to the Rx, with $l_{\text{Tx}} = \lVert \mathbf{r}_{\text{Tx}} - \mathbf{r}_{\text{Rx}} \rVert$ denoting the length of the corresponding path, $\lVert . \rVert$ representing the $l_2$ norm of the argument, and $P_t$, $G_t$, and $G_r$ denoting the transmit signal amplitude, the gain of the transmit antenna and the gain of the receiver antenna respectively. Furthermore, $\alpha_{g}$ and $l_{g}$ are the amplitude and length of the path reflected off of the ground, $\lambda$ is the wavelength of the signal, and $\eta(\mathbf{r}_{\text{Rx}})$ is the signal noise at the Rx.  We then have $\alpha_{g} = \frac{\lambda P_t G_t G_r \gamma_g}{4\pi l_{g}}$, where $\gamma_g = \frac{\sin \theta-Z}{\sin \theta+Z}$, $Z = \frac{\sqrt{\epsilon_r - \cos^2 \theta}}{\epsilon_r}$, $\theta$ is the angle between the ground plane and the signal ray bouncing off the ground, and $\epsilon_r$ is the relative permittivity of the ground \cite{goldsmith2005wireless}. Fig.~\ref{fig_ground_params} shows the parameters of the ground reflection.   

Next, let $n \in \{1, 2, \dots, N\}$ denote the indexed set of the other paths arriving at the receiver, excluding the direct path from the Tx to the Rx and the ground reflection path.  \revv{Let $\mathbf{r}_n$ represent} the location of the last object that the $n^{\text{th}}$ path visits on its way to the receiver, as shown in Fig.~\ref{fig_general_scenario}. $R_n$ then denotes the attenuation that the $n^{\text{th}}$ path experiences from the Tx all the way up to $\mathbf{r}_n$. More specifically, this attenuation is the result of the distance traveled from the transmitter all the way up to $\mathbf{r}_n$, the potential reflections from/attenuation by other objects that the $n^{\text{th}}$ path encounters before arriving at $\mathbf{r}_n$, as well as the attenuation experienced through interaction with the object at $\mathbf{r}_n$. We then have $\alpha_n = \frac{\lambda P_t G_t G_r R_n}{4 \pi \lVert \mathbf{r}_n - \mathbf{r}_{\text{Rx}}\rVert}$. Note that in the special case that \revv{the only propagation phenomena} prevalent in the environment is primary reflections off of objects in the area, i.e., the object at $\mathbf{r}_n$ is the only object the $n^{\text{th}}$ path visits before arriving at the receiver, we will have $R_n = \frac{\gamma}{4\pi \lVert \mathbf{r}_{\text{Tx}} - \mathbf{r}_n \rVert}$, where $\gamma$ is the reflection coefficient of the \revv{object} at the point of reflection.   

In this paper, we are interested in predicting the wireless channel of Eq.~\ref{eq_point_signal}, its variations, and the detailed makeup of the rays arriving at any such given location in the area, using a \revv{small number of} prior channel power measurements in the area.  We show how we can methodically design the locations of the prior measurements, in order to extract the most vital information about the rays propagating through the area, hence resulting in a high-quality channel prediction. More specifically, we utilize the mobility of an unmanned vehicle and show how to design its route in order to optimize the location/configuration of the prior measurement collection phase. Along its route, the unmanned vehicle will form antenna arrays, which we then utilize for extracting information from the \revv{prior channel} measurements.
We next summarize the received signal power model across an antenna array in order to set the stage for our proposed approach of Sec.~\ref{sec_proposed_framework}. 

Consider the Rx antenna array of Fig.~\ref{fig_antenna_array}, where the antenna array is generated by a robot that is moving along its route. Let $\mathbf{r}_a$ denote the location of the first antenna. The complex baseband received signal $c(d)$ at the Rx array can then be written as a function of the distance $d$ along the array as follows \cite{goldsmith2005wireless,karanam2018magnitude}:
\vspace{-0.05in}
\begin{align}\label{eq_complex_baseband}
c(d) = \alpha_{\text{Tx}} & e^{j \left(\frac{2\pi}{\lambda}l_{\text{Tx}} - \frac{2\pi}{\lambda}d\cos\phi_{\text{Tx}} \right)} + \alpha_g e^{j \frac{2\pi}{\lambda} l_g(d)}   \nonumber \\
 & + \sum_{n=1}^N \alpha_n e^{j \left(\frac{2\pi}{\lambda}l_n - \frac{2\pi}{\lambda}d\cos\phi_n \right)} + \eta(d),
 \vspace{-0.1in}
\end{align}
where $\alpha_\text{Tx}$ and $l_{\text{Tx}}$ are the amplitude and length of the direct path from the Tx to $\mathbf{r}_a$, $\alpha_g$ \rev{is the amplitude of the ground path at $\mathbf{r}_a$}, $l_g(d)$ is the length of the ground path measured across the antenna array, $\alpha_n$ and $l_n$ are the amplitude and length of the $n^{\text{th}}$ path (excluding direct and ground paths) propagating from the Tx, interacting with the objects in the area and arriving at $\mathbf{r}_a$, as discussed in Eq.~\ref{eq_point_signal}.\footnote{\rev{Note that $l_g$ is written} as a function of distance $d$ along the array, since the length of the ground path varies across the antenna array.} Furthermore, $\phi_{\text{Tx}}$ and $\phi_n$ are the angles-of-arrival~(AoA) of the direct path and the $n^{\text{th}}$ object-path, with respect to the antenna array, respectively. Without loss of generality, angles $\phi_{\text{Tx}}$ and $\phi_n$ are \revv{are shown in a 2D plane} in Fig.~\ref{fig_antenna_array}. As can be seen from Eq.~\ref{eq_complex_baseband}, these AoAs characterize the progression of the phase of the corresponding signal path across the antenna array, with respect to the phase at the first antenna in the array. 
Note that we form short antenna arrays along the robot route in a far-field setting. Thus, the angles and the amplitudes corresponding to the rays discussed in Eq.~\ref{eq_complex_baseband} \revv{can be assumed approximately constant} over the considered antenna array.

The wireless received signal power along the array can then be derived from Eq.~\ref{eq_complex_baseband} as,
\vspace{-0.1in}
\begin{align}\label{eq_channel_power}
|c(&d)|^2 \; \approx \;\;\; \alpha_{\text{Tx}}^2 + \alpha_g^2  +  \sum_{n=1}^N \alpha_n^2 +  \nonumber \\
& \alpha_{\text{Tx}} \alpha_g \left\{ e^{j\frac{2\pi}{\lambda} [l_{\text{Tx}} - d\cos \phi_{\text{Tx}} -l_g(d)]} + e^{-j\frac{2\pi}{\lambda} [l_{\text{Tx}} - d\cos \phi_{\text{Tx}} -l_g(d)]} \right\} \nonumber \\
& + \sum_{n=1}^N \alpha_{\text{Tx}} \alpha_n \left\{ e^{j\frac{2\pi}{\lambda} [(l_{\text{Tx}} - l_n) - d(\cos \phi_{\text{Tx}} - \cos \phi_n) ]} + \right.  \nonumber \\ & \left. \;\;\;\;\;\;\;\;\;\;\;\;\;\;\;\;\;\;\;\;\;\; e^{-j\frac{2\pi}{\lambda} [(l_{\text{Tx}} - l_n) - d(\cos \phi_{\text{Tx}} - \cos \phi_n) ]} \right\},
\end{align}
where we neglect the weaker terms corresponding to the cross terms between various reflected paths, since the direct path from the Tx is dominant as compared to the reflected paths \cite{karanam2018magnitude, karanam2019tracking}. The key parameters of the rays are then as follows, based on Eq.~\ref{eq_channel_power}:
\vspace{-0.05in}
\begin{align}\label{eq_params}
\;\;\text{Path gains:}& \;\;\; \alpha_{\text{Tx}}, \alpha_g, \alpha_n \;\;\forall \; n = 1, \dots, N  \nonumber \\
\;\;\text{Path lengths:}& \;\;\; l_{\text{Tx}}, l_g, l_n \;\; \forall \; n = 1, \dots, N  \\
\;\;\text{Angles of arrival:}& \;\;\; \phi_{\text{Tx}}, \phi_n \;\; \forall \; n = 1, \dots, N. \nonumber
\end{align}

In this paper, we are interested in the problem of robotic channel prediction, where a robot first collects a few prior signal power measurements in the area, for the purpose of channel prediction. Without any phase synchronization between the Tx and Rx (which is typically the case with off-the-shelf devices), it is only possible to measure the received signal power (i.e., Eq.~\ref{eq_channel_power}) reliably at the Rx, due to frequency and timing offsets between \revv{the Tx and Rx} \cite{zhuo2016identifying}. As a result, we are interested in 1) extracting the parameters of Eq.~\ref{eq_params} from the signal power measurements across the antenna arrays formed by the robot's path, \revv{and} 2)~designing the path of the robot such that the extraction of the parameters of Eq.~\ref{eq_params} across its path allows it to fully predict the detailed ray makeup at any unvisited location in the space. 
In the next section, we propose a new framework that enables us to achieve these goals. 

\vspace{-0.1in}
\section{Proposed Framework}\label{sec_proposed_framework}
So far, we have discussed the signal model and the ray parameters that are necessary for fully characterizing the wireless received signal at any location. \rev{In this section, we are interested in predicting the makeup of the rays at any unvisited location}. However, predicting the \rev{detailed} ray parameters of Eq.~\ref{eq_params} at unvisited locations, without any knowledge of the \revv{geometry} and material properties of the objects in the area, is a very challenging unsolved problem, as discussed in Sec.~\ref{sec_intro}. 
We thus propose a new framework, where we focus on the rays passing through the area of interest, and show how to extend the rays along their paths in order to characterize the wireless channel at unvisited locations. More specifically, we are interested in 1) extracting the fundamental parameters of the rays that arrive at the prior measurement locations in the area, 2) extending these rays along their paths to other locations of interest in the area, and finally 3) predicting the key parameters of the rays at these unvisited locations. 

In order to optimally extract the key parameters of all the rays passing through the region of interest, we propose an enclosure-based framework to collect prior wireless measurements along the boundary of the region where we are interested in predicting the wireless channel. The key intuition for enclosing the region of interest is that we are intercepting all the rays that pass through the workspace, since each ray that passes through the region of interest intersects the boundary at a minimum of two locations. In this paper, we utilize these intersections to estimate the corresponding fundamental parameters of the rays at the boundary, and then show how to extend the rays through the region of interest to fully predict the detailed ray \revv{makeup} at unvisited locations. We next discuss our proposed approach in more detail.

Consider the scenario shown in Fig.~\ref{fig_prediction_scenario}. In order to predict the channel quality at any location in the empty region $\mathcal{A}$, we propose to collect the prior wireless channel power measurements along the boundary of the region, using a robot, as shown in Fig.~\ref{fig_prediction_scenario}. We note that if the region of interest is not empty and contains an object, one can easily re-draw the boundaries in the area to form multiple empty regions that do not contain any objects within them and apply the method we shall propose to each such region. 
We further note that since most robots have on-board vision systems for obstacle avoidance, \rev{they can easily form the empty regions accordingly.
Finally}, while we drew a \rev{sample} rectangular prediction region in Fig.~\ref{fig_prediction_scenario}, we note that the prediction area does not have to be rectangular or \revv{even convex}. For a non-convex prediction region, a ray can intersect the boundary at more than 2 points. In such cases, we only consider the two points, \rev{one on each side}, that are closest to the prediction point, along the ray passing through the point, and use our proposed approach on those \revv{prior} samples. In this manner, our proposed approach can \revv{be easily} extended to regions that are non-convex.

\begin{remark}
\revv{Fig.~\ref{fig_prediction_scenario} shows a case where the enclosure and the corresponding prediction region lie in 2D. This is due to the fact that, in practice, common off-the-shelf RF antennas typically have a limited beamwidth in the elevation dimension, which results in the receiver antenna mainly receiving the rays close to the horizontal plane passing through the antenna. Thus, in such cases, a 2D prediction region is a good approximation to the real-world setup (our experimental tests of the next section also confirm this). However, we note that our proposed approach is easily extendable to 3D scenarios, where, our enclosure-based framework implies that we would form a 3D enclosure of the prediction region and collect prior measurements along the boundary in 3D. Collecting measurements along a 3D enclosure can be implemented, for instance, by using an unmanned aerial vehicle that moves around the prediction region. We then use our proposed approach on such measurements in 3D, in order to extract and predict the detailed parameters of the rays in the area.}
\end{remark}

In the rest of this section, we first discuss how to extract the key ray parameters at the boundary points and then show how we can predict the parameters at unvisited locations, using the prior measurements at the boundary.

\begin{figure}
    \centering
    \includegraphics[width=0.9\linewidth]{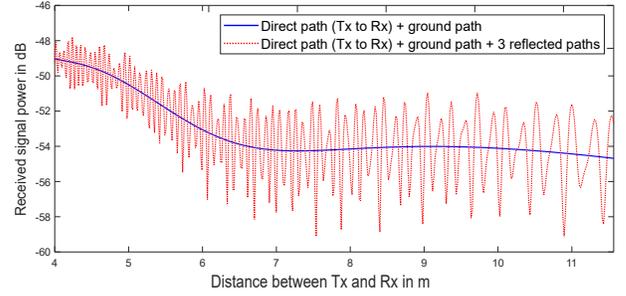}
    \caption{A sample simulation showing the contributions of different terms in Eq.~\ref{eq_channel_power} to the variations in the received signal power along a route.} 
    \vspace{-0.2in}
    \label{fig_signal_trends}
\end{figure}

\vspace{-0.15in}
\subsection{Estimating Ray Parameters at a Boundary Antenna Array}\label{sec_estimating_params}
Consider the signal power measurements $|c(d)|^2$ across an antenna array, as discussed in Eq.~\ref{eq_channel_power}. 
An important observation to note in the model discussed in Eq.~\ref{eq_channel_power} is that the first \revv{three terms mainly} contribute to the mean of the received power.\footnote{\revv{We note that $\sum_{n=1}^N \alpha_n^2$ also contributes to the mean of the signal. However, it is typically much weaker than $\alpha_{\text{Tx}}^2$ and as such it is not considered in the discussion on the mean.}} On the other hand, the rapid variations in the signal mainly arise from the last summation term in the equation, corresponding to the rays interacting with the objects in the area. To visualize this in more detail, consider the trends shown in Fig.~\ref{fig_signal_trends}.
As can be seen from the figure, different terms in the equation contribute to different trends in the signal power. For instance, \rev{we know from Eq.~\ref{eq_point_signal} that} on a route where the distance between the Tx and the Rx is increasing, $\alpha_{\text{Tx}}^2$ \rev{(i.e., the power of the direct path from Tx to Rx)} is monotonically decreasing along the route. Upon adding the terms corresponding to the ground path, we can see \rev{from the figure} that it then results in a slowly-oscillating mean of the signal power across the route. Finally, the rapid variations in the signal then arise from the reflections off of different objects in the area. 
Thus the figure intuitively shows that the combination of the direct path and the ground reflection contributes to the mean of the signal power, and that the term $\sum_{n=1}^N \alpha_n^2$ does not result in any noticeable change in the mean \revv{(see Footnote~2)}.
A more \rev{detailed} mathematical characterization \rev{of the ground path frequency content} is provided in Appendix~\ref{app_ground_path}. We next \revv{show how to} utilize these different trends in order to extract the desired parameters at an antenna array on the boundary measurements. 
\begin{itemize}
    \item \textbf{Estimating} $\bm{l}_{\text{Tx}}, \bm{l}_g, \bm{\phi}_{\text{Tx}}$: Since we know the location of the Tx and the Rx, these parameters are straightforward to estimate. The length of the ground path between a Tx and a Rx antenna can be calculated as $l_g = 2\sqrt{(l_{\text{Tx}}/2)^2 + h_a^2}$, where $h_a$ is the height of the antennas above the ground.\footnote{Note that this expression for $l_g$ is for the scenario where both the Tx and Rx antennas are located at the same height above the ground. Our proposed framework can equivalently be extended to the case with different antenna heights.} $\phi_{\text{Tx}}$ can also be similarly calculated, as shown in Fig.~\ref{fig_antenna_array}. 
    
    \item \textbf{Estimating} $\bm{\alpha}_{\text{Tx}}, \bm{\alpha}_g$: As discussed in Fig.~\ref{fig_signal_trends}, the signal power corresponding to the term \revv{$|c_0(\bm{r}_{\text{Rx}})|^2 \approx \alpha_{\text{Tx}}^2 + \alpha_g^2 + \alpha_{\text{Tx}} \alpha_g \left( e^{j\frac{2\pi}{\lambda} [l_{\text{Tx}} -l_g]} + e^{-j\frac{2\pi}{\lambda} [l_{\text{Tx}} -l_g]} \right)$ contributes to the slowly-oscillating mean of the measurements along a robot route}, where $\bm{r}_{\text{Rx}}$ is the location of the receiver antenna at a sample point on the route. Note that we express the signal power here as a function of $\bm{r}_{\text{Rx}}$ and not the distance $d$ along the route, because this component of the overall signal power only depends on the locations of the Tx and Rx antennas. Thus, by knowing the locations and heights of the Tx and Rx antennas, the only remaining unknowns in the above expression are the ground relative permittivity $\epsilon_r$ and the product $G = P_t G_t G_r$. Suppose that we denote the set of all measurement points on the prior boundary route as $\mathcal{R}_b$. We then solve for $\epsilon_r$ and $G$, by minimizing the mean squared error between the measured signal power and the theoretical mean, \revv{at all the points on the prior measurement route $\mathcal{R}_b$}, as follows, 
    \vspace{-0.15in}
    \begin{align}
        \hat{\epsilon}_r\; , \; \hat{G} \; = &\; \arg \min_{\epsilon_r, G} \frac{1}{M} \sum_{\bm{r}_b \in \mathcal{R}_b} \left(10 \log_{10} |\hat{c}(\mathbf{r}_b)|^2 \right. \nonumber \\ & \left. - 10 \log_{10} |c_0(\mathbf{r}_b, \epsilon_r, G)|^2\right)^2,
    \end{align}
    where $M$ denotes the number of measurement points on the prior boundary route, \revv{$|\hat{c}(\mathbf{r}_b)|^2$ is the measured power at a sample Rx location on the} prior measurement route, and $|c_0(\mathbf{r}_b, \epsilon_r, G)|^2$ is the theoretical mean \revv{at that location on the route,} for a given $\epsilon_r$ and $G$. 
    Upon estimating these parameters, we can then calculate the value of $\alpha_{\text{Tx}}$ and $\alpha_g$ at \revv{any point on the prior measurement route} as well as at any other unvisited location in the area, \rev{using} $\alpha_{\text{Tx}} = \frac{\lambda \hat{G}}{4\pi l_{\text{Tx}}}$ and $\alpha_{g} = \frac{\lambda \hat{G} \gamma_g}{4\pi l_{g}}$, where $\gamma_g = \frac{\sin \theta - Z}{\sin \theta + Z}$, and $Z = \frac{\sqrt{\hat{\epsilon}_r - \cos^2 \theta}}{\hat{\epsilon}_r}$. Thus, in this manner, we utilize the prior wireless measurements collected along the boundary routes to estimate the ground and antenna properties, and thereby the \rev{direct} Tx \revv{path} and ground path amplitudes. 
    
    \item \textbf{Estimating} $\bm{\phi}_n$: Estimating $\phi_n$ is traditionally a generic angle-of-arrival estimation problem at an antenna array, with many established solutions \cite{van1988beamforming,schmidt1986multiple,roy1989esprit}. However, these approaches typically require signal phase measurements at the antenna array, which is not always available in a reliable manner \cite{zhuo2016identifying}. As a result, in this paper, we instead use the approach discussed in \cite{karanam2018magnitude, karanam2019tracking} to estimate the AoA of the signal paths using only the signal power measurements at an antenna array. More specifically, as discussed in \cite{karanam2018magnitude, karanam2019tracking}, by performing an FFT on a \revv{small} array of $|c(d)|^2$ measurements, we obtain a Fourier spectrum as follows:
    \vspace{-0.1in}
    \begin{align}\label{eq_fourier_spectrum}
        \mathcal{C}(f) = \sum_{n=1}^N &\alpha_{\text{Tx}} \alpha_n  \left\{ e^{j \frac{2\pi}{\lambda} (l_{\text{Tx}} - l_n)} \delta \left( f - \frac{\cos \phi_{\text{Tx}} - \cos \phi_n}{\lambda}\right) \right. \nonumber \\ & \left. + e^{-j \frac{2\pi}{\lambda} (l_{\text{Tx}} - l_n)} \delta \left( f + \frac{\cos \phi_{\text{Tx}} - \cos \phi_n}{\lambda}\right) \right\}. 
    \end{align}
    \vspace{-0.05in}
    Note that in this equation, we have omitted the \rev{impact of the first four terms of Eq.~\ref{eq_channel_power}}, since we can remove their contribution by filtering out the content corresponding to the zero frequency and its neighboring bins in the FFT (i.e., \rev{the low-frequency content}). We then see peaks in the Fourier spectrum at $\lambda$-normalized frequencies $\pm (\cos \phi_{\text{Tx}} - \cos \phi_n)$ for every path $n \in [1, N]$. Without loss of generality, suppose that we denote $\psi_n = \cos \phi_{\text{Tx}} - \cos \phi_n$. Thus, \rev{by} using only the signal power, we can estimate the value $|\psi_n|$ for each path in the area. Furthermore, we can calculate the AoA of the $n^{\text{th}}$ path to be $\cos^{-1} (\cos \phi_{\text{Tx}} \pm |\psi_n|)$. While this does not uniquely characterize $\phi_n$, in the next section, we shall discuss how this information at the boundary array is indeed sufficient to predict the detailed ray parameters at any location inside the region $\mathcal{A}$. 
    \item \textbf{Estimating} $\bm{\alpha}_n$: This parameter contains information about the strength or power of the $n^{\text{th}}$ path at the antenna array. We estimate this parameter through the Fourier spectrum, similar to $|\psi_n|$. More specifically, while the location of the peak in the spectrum contains information about the AoA of the $n^{\text{th}}$ path \rev{($\phi_n$)}, the absolute magnitude of the same peak contains information about the amplitude of the \rev{corresponding} path. As can be seen from Eq.~\ref{eq_fourier_spectrum}, the absolute value of the peak corresponding to the $n^{\text{th}}$ path is $\alpha_{\text{Tx}} \alpha_n$. Since we have already estimated $\alpha_{\text{Tx}}$ at the antenna array, we can thus compensate for it and subsequently estimate $\alpha_n$, for all the \revv{paths arriving at the array}. In the next section, we show how to use this information to predict the path amplitude at any unvisited location within the region $\mathcal{A}$, without the need for \rev{estimating the properties of} the objects that generated the corresponding path.
    \item \textbf{Estimating} $\bm{l}_n$: The parameter $l_n$ is the total length of the $n^{\text{th}}$ path, starting from the Tx, propagating through the area, undergoing any number of reflection(s)/diffraction(s), and ultimately arriving at the first antenna in the array. The estimation of $l_n$ is very challenging, especially when the locations of the objects that interacted with the rays are not known. We observe that the information about $l_n$ exists in the Fourier spectrum discussed in Eq.~\ref{eq_fourier_spectrum}, in the complex phase of the peak corresponding to the $n^{\text{th}}$ path. More specifically, the complex phase of the $n^{\text{th}}$ peak is $e^{\pm j\frac{2\pi}{\lambda} (l_{\text{Tx}} - l_n)}$. Since we know the value of $l_{\text{Tx}}$, we can thus estimate two possibilities for the value of $e^{j \frac{2\pi}{\lambda}l_n}$. In the next section, we show how we can use this information to predict the path length at any unvisited location inside the region $\mathcal{A}$.  
\end{itemize}

\vspace{-0.15in}
\subsection{Predicting \rev{the} Ray Parameters at an Unvisited Location}\label{sec_predicting_params}
So far we have discussed how we can estimate the fundamental parameters of the rays arriving at any antenna array on the boundary of the region of interest, albeit with an ambiguity in the AoA and the complex phase corresponding to the length of each path. We next discuss how we utilize the information we have extracted from the boundary routes to predict the detailed ray parameters in the interior of the region. 

Consider the scenario shown in Fig.~\ref{fig_prediction_scenario}. Suppose we want to predict the channel at a sample point $p$. As discussed in Sec.~\ref{sec_estimating_params}, the parameters $l_{\text{Tx},p}, l_{g,p}, \phi_{\text{Tx},p}$ \fin{(corresponding to the direct and ground paths)} are already known based on the location of point $p$, where the ``$p$" in the subscripts of the parameters denotes the parameter value at the prediction point $p$. We can also estimate $\alpha_{\text{Tx},p}$ and $\alpha_{g,p}$ using the parameters estimated from the boundary measurements, as discussed in the previous section. Next, in order to fully predict the makeup of the rays and the channel at this location, we need to first estimate which rays in the area pass through $p$, and then estimate the parameters of those rays at the prediction point. \textbf{The key principle here is that we consider this prediction problem from the perspective of the point where we want to predict the parameters}. More specifically, we draw rays passing through the prediction point $p$, at angles ranging from $0$ to $2\pi$, in order to scan the angular space and check for valid rays that arrive at this point. As a result, every such ray then intersects the boundary measurement routes at two points \rev{in} Fig.~\ref{fig_prediction_scenario} \fin{(see the discussion in Sec.~\ref{sec_estimating_params} for handling non-convex areas)}. Consider one such sample candidate ray that passes through the prediction point, at angle $\phi_c$, as shown in the figure. 
We then form antenna arrays at the two intersections \rev{along the boundary}.
In order to predict the ray parameters at the point $p$, we first estimate the ray parameters at the two intersecting arrays, i.e., we first estimate the AoAs, path amplitudes, and complex phases of the rays arriving at these two boundary arrays. Next, we check if there exists any ray at the two arrays that matches the direction of the candidate ray at angle $\phi_c$ that we drew through the prediction point. We perform this check by calculating the values of $|\psi_{c,1}| = |\cos \phi_{\text{Tx},1} - \cos \phi_c|$ and $|\psi_{c,2}| = |\cos \phi_{\text{Tx},2} - \cos \phi_c|$, where $\phi_{\text{Tx,1}}$ and $\phi_{\text{Tx,2}}$ are the estimated $\phi_{\text{Tx}}$ at the two corresponding boundary antenna arrays.
If $(|\psi_{c,1}|, |\psi_{c,2}|)$ belong to the set of peaks estimated from the spectrum at \rev{the} corresponding boundary arrays, we then declare the ray drawn at angle $\phi_c$ \rev{as} a valid ray that indeed passes through the area. Furthermore, we now know the AoA of this ray to be $\phi_c$, thus resolving the ambiguity that arose when we estimated the AoAs at the boundary arrays using only the signal power measurements. 

\begin{remark} \label{rem_AoA_ambiguity}
Note that at any antenna array (even when \fin{the} signal phase is available), there always exists an array half-space ambiguity, where we cannot uniquely estimate the AoA of the ray from the two possibilities arising due to the two half-spaces created by the antenna array. More specifically, two rays at angles $\phi$ and $2\pi - \phi$, with respect to an antenna array, result in the exact same parameter $\psi$ estimated at the array, since $\cos \phi = \cos (2\pi - \phi)$. As a result, given a signal measurement at an antenna array, there exist two possibilities for the AoA of a ray corresponding to the two half-spaces. In addition, as we showed in the previous section, when we estimate the AoA at an array with only signal power measurements, there is another ambiguity regarding \rev{the} side of the Tx \rev{from which} the ray is coming (the estimated angle is $\cos^{-1} (\cos \phi_{\text{Tx}} \pm |\psi_n|)$).
However, in our proposed prediction framework, we check the validity of a candidate ray at two separate antenna arrays on the boundary measurements, thus implicitly resolving \rev{both} ambiguities by utilizing the diversity of the boundary measurement arrays. 
\end{remark}

\begin{remark}
As discussed so far, we use two arrays on the boundary to resolve the ambiguity regarding the AoA of a ray that passes through the region. While our approach is successful in eliminating the ambiguity \rev{corresponding to the} four possible AoA solutions \rev{at an array}, there is a \rev{very small} probability \rev{that there exists} complementary rays at both boundary antenna arrays, at specific angles such that they both correspond to the same $(|\psi_{c,1}|, |\psi_{c,2}|)$ pair, \rev{thus resulting} in a false positive declaration of a valid ray at $\phi_c$ when there does not exist a ray at that angle. However, such scenarios are \fin{rare}. 
\end{remark}

Next, we need to predict the amplitude of this ray at point $p$. In order to do so, we first estimate the \rev{corresponding} path amplitudes at the two arrays where this candidate ray intersects the boundary, using the approach discussed in Sec.~\ref{sec_estimating_params}. Suppose we denote these amplitudes as $\alpha_{c, 1}$ and $\alpha_{c, 2}$, where the ray intersects point 1 first, passes through $p$ and then intersects point 2, as shown in Fig.~\ref{fig_prediction_scenario}. \fin{Note that} since we have drawn the candidate ray at angle $\phi_c$, it is easy to geometrically calculate the intersection points on the boundary, \fin{as well as} the order in which the ray intersects those points. Based on the model for $\alpha_n$ discussed in Sec.~\ref{sec_estimating_params}, we have $\alpha_{c, 1} = \frac{\lambda P_t G_t G_r R_c}{4\pi \lVert \mathbf{r}_1 - \mathbf{r}_c \rVert}$ and $\alpha_{c, 2} = \frac{\lambda P_t G_t G_r R_c}{4\pi \lVert \mathbf{r}_2 - \mathbf{r}_c \rVert}$, where $\mathbf{r}_c$ denotes the location of the last object that the candidate ray visits on its way to the receiver, and $\mathbf{r}_1$ and $\mathbf{r}_2$ denote the locations of points 1 and 2 on the boundary respectively. After some \rev{straight-forward} derivations, we can show that the path amplitude $\alpha_{c, p}$ at the prediction point can be written as follows,
\vspace{-0.05in}
\begin{equation}\label{eq_alpha_estimation} 
    \alpha_{c, p} = \frac{\alpha_{c, 1} \alpha_{c, 2} \lVert \mathbf{r}_1 - \mathbf{r}_2 \rVert} { \alpha_{c, 1} \lVert \mathbf{r}_1 - \mathbf{r}_p \rVert + \alpha_{c, 2} \lVert \mathbf{r}_2 - \mathbf{r}_p \rVert }.
\end{equation}
Thus, our proposed route design and prediction framework enables us to derive the path amplitude at an unvisited location without the need for \fin{estimating the material property/geometry of} the objects that interact with the signal rays. 

Finally, we need to estimate the phase term $e^{j\frac{2\pi}{\lambda} l_{c, p}}$ of the ray at the prediction point.\footnote{Note that in order to reconstruct and predict the channel according to Eq.~\ref{eq_complex_baseband}, we only need to estimate/predict the quantity $e^{j\frac{2\pi}{\lambda} l_{c, p}}$ and not the exact value of $l_{c,p}$.} As discussed in Sec.~\ref{sec_estimating_params}, we can estimate two possible solutions for the complex phase of a ray at any array on the boundary. However, since we have resolved the ambiguity in the AoA for the candidate ray, we can subsequently calculate $\psi_c = \cos \phi_{\text{Tx}} - \cos \phi_c$ without any ambiguity in its sign. We can then resolve the ambiguity in the path length as well by estimating the complex phase of only the peak that corresponds to $\psi_1 = \cos \phi_{\text{Tx, 1}} - \cos \phi_c$ in the spectrum at the first boundary array. Suppose that this complex phase at point 1 on the boundary is denoted by $\mu_{c, 1} = \frac{2\pi}{\lambda} (l_{\text{Tx}, 1} - l_{c, 1})$, where we have already estimated the value of $l_{\text{Tx}, 1}$, and $l_{c,1}$ is the distance from the last object along the path of the ray to point 1 on the boundary. Thus, we can then calculate the complex phase at point $p$ as follows:
\begin{equation}\label{eq_complex_phase}
    e^{j\frac{2\pi}{\lambda}l_{c, p}} = e^{j \mu_{c, 1}} \times e^{-j \frac{2\pi}{\lambda} l_{\text{Tx}, 1}} \times e^{-j\frac{2\pi}{\lambda} \lVert \mathbf{r}_1 - \mathbf{r}_p\rVert},
\end{equation}
where we have compensated for the impact of the Tx path length, and added the extra distance that the ray traveled from point 1 on the boundary to the prediction point $p$. 

\rev{Overall}, we have shown how to estimate all the parameters for a \rev{valid} candidate ray at angle $\phi_c$, at the prediction point $p$. We then repeat the same process for all \rev{the} possible candidate rays with angles ranging from $0$ to $2\pi$. As a result, we can estimate all the parameters that constitute the makeup of the rays at the prediction point $p$. We then use Eq.~\ref{eq_point_signal} to reconstruct the complex baseband received signal, at \rev{the prediction point}, \fin{and further predict the resulting received signal power}. We subsequently repeat this procedure to predict the received signal at any other unvisited location in the region of interest.

\vspace{-0.15in}
\section{Experimental Results}\label{sec_exp_results}
In this section, we validate our proposed approach for channel prediction through extensive experiments in multiple areas. We first discuss our experimental setup for collecting measurements. Next, we validate our proposed approach for predicting the detailed parameters of the rays by comparing the predicted and true path amplitudes and AoAs corresponding to objects in the area. We then discuss our experimental results in more complex scenarios where several objects in the area interact with the wireless signals. We finally compare our approach to the state-of-the-art in channel prediction. 

\vspace{-0.15in}
\subsection{Experimental Setup}
In our experiments, we collect WiFi signal magnitude measurements using a robot moving in the area of interest. More specifically, for the Tx, we use a USRP N210 Software Defined Radio (SDR) \cite{web_ettus} operating at the first WiFi sub-channel at 2.4 GHz. For the Rx, we mount another USRP N210 SDR on a Pioneer 3-AT ground robot \cite{web_pioneer} that can move around the area of interest and collect wireless signal power measurements. We operate the SDR in a narrowband WiFi channel at 2.4 GHz. The beamwidth of the Rx antennas along the elevation angle is \fin{$20^\circ$ above and below the horizontal plane}, which results in the rays mainly arriving at the Rx antenna in a 2D horizontal plane. Thus, our 2D enclosure-based method can be applied.

\rev{In} order to extract the path amplitudes and angles at the boundary measurement arrays, as discussed in Eq.~\ref{eq_fourier_spectrum}, we use a \rev{1-meter-long moving} array of measurements \rev{on the robot route} to calculate the spectrum. In order to detect peaks in the spectrum, we use a threshold of $\beta_{\text{th}} \times \max (|\mathcal{C}_p|)$ over the window of measurements. We set $\beta_{\text{th}} = 0.15$ for all the experiments. Any peak above this threshold is considered to be due to a signal path arriving at the antenna array. We next discuss our experimental results and validate various aspects of our proposed approach. 

\begin{figure}
    \centering
    \includegraphics[width=0.85\linewidth]{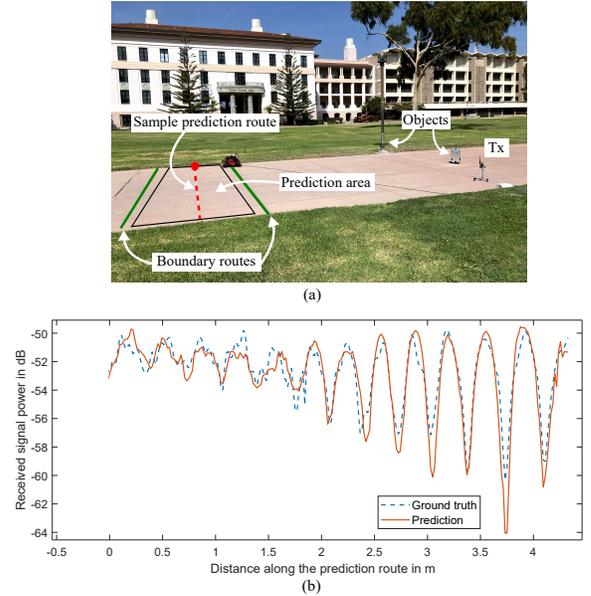}
    \vspace{-0.1in}
    \caption{(a) Area \#1 -- A controlled experiment area with a Tx and two objects. The green solid lines represent the boundary measurement routes. The prediction area is 5~m $\times$ 2~m, and marked in the figure. (b) Channel prediction along a sample prediction route marked in (a), \fin{starting at the filled circle marked on the route}. The blue dashed curve is the ground-truth received signal power measurements along the route of interest, while the red solid curve is the signal power predicted by our approach on the same route.}
    \vspace{-0.1in}
    \label{fig_phy_area_result}
\end{figure}

\subsection{Channel Prediction in a Controlled Environment (Area \#1)}
We start by testing our approach in a more controlled environment where there are only a few objects present. More specifically, consider Area \#1, shown in Fig.~\ref{fig_phy_area_result}~(a). As can be seen from the figure, there is one Tx and two objects in the area. The robot collects its prior wireless signal power measurements along the \rev{marked} boundary routes, and uses these measurements to predict the makeup of the rays elsewhere in the area. The prediction area is marked in the figure, and is of dimensions 5~m~$\times$~2~m.

\begin{figure}
    \centering
    \includegraphics[width=0.9\linewidth]{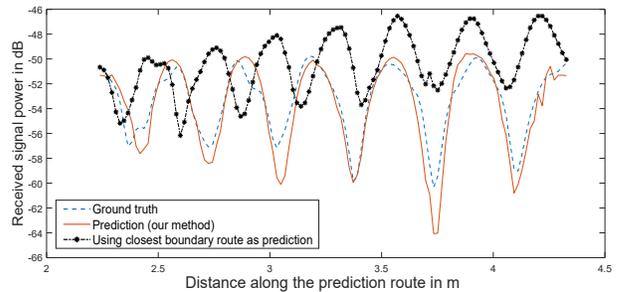}
    \vspace{-0.15in}
    \caption{Received signal power comparison between measurements on the prediction route and the closest boundary route shown in Fig.~\ref{fig_phy_area_result}~(a). For easy visualization, the plot shows a zoomed-in version of the second half of the result discussed in Fig~\ref{fig_phy_area_result}~(b).} 
    \label{fig_phy_area_comparison}
    \vspace{-0.05in}
\end{figure}

\begin{figure}
    \centering
    \includegraphics[width=1\linewidth]{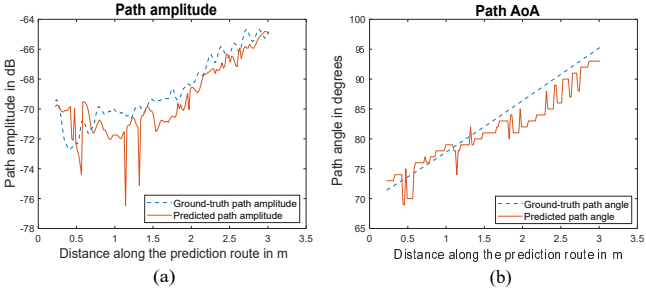}
    \vspace{-0.2in}
    \caption{Predicting the detailed makeup of a path reflected from a sample object in the environment, \rev{along the sample prediction route of Fig.~\ref{fig_phy_area_result}~(a)} -- (a) Path amplitude \rev{prediction}, and (b) Angle of arrival prediction.}
    \vspace{-0.1in}
    \label{fig_phy_area_params}
\end{figure}

Fig.~\ref{fig_phy_area_result}~(b) shows the measured and predicted signal power along a sample prediction route, marked on the figure. As can be seen, our proposed approach can accurately predict the channel and match the detailed signal variations that arise from reflections off of the objects in the area. A simple benchmark approach here would have been to use the received signal power measurements from the closest parallel boundary route as a prediction for the signal power on the prediction route. However, as can be seen from Fig.~\ref{fig_phy_area_comparison}, measurements from a nearby boundary route that is only 1 m away from the prediction route are far from the true channel value on the prediction \rev{route}. 

Fig.~\ref{fig_phy_area_params}~(a) next compares the ground-truth and predicted path amplitude, along the prediction route, for a sample path reflected from a sample object in the environment. As can be seen, our approach can well predict the varying path amplitude along the route, without the need for localizing the object. Fig.~\ref{fig_phy_area_params}~(b) further shows the comparison between the corresponding AoAs, which shows a very good match.

Overall, we can see that our approach accurately predicts the underlying parameters of the rays passing through unvisited points in this area.

\begin{figure}
    \centering
    \includegraphics[width=0.75\linewidth]{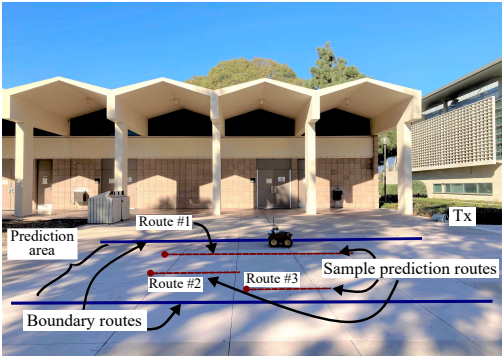}
    \vspace{-0.1in}
    \caption{Area \#2 of dimensions 8 m $\times$ 3.5 m -- boundary routes and sample prediction routes are marked.}
    \vspace{-0.1in}
    \label{fig_chem_area}
\end{figure}

\vspace{-0.1in}
\subsection{Channel Prediction in More Complex Areas}
\begin{figure}
    \centering
    \includegraphics[width=0.95\linewidth]{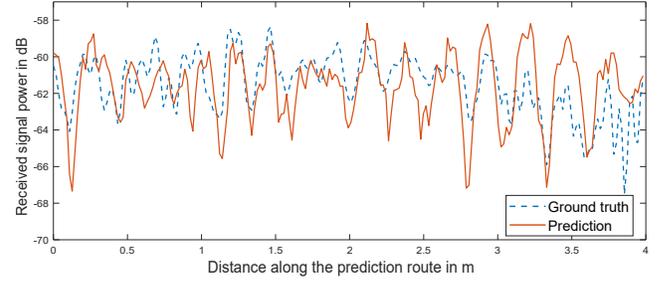}
    \vspace{-0.1in}
    \caption{Channel prediction in Area \#2, using our proposed framework. The plot shows the comparison between the ground truth and predicted signal power measurements along prediction route \#1 of Fig.~\ref{fig_chem_area}.}
    \label{fig_chem_area_result2}
\end{figure}

\begin{figure}
    \centering
    \includegraphics[width=1\linewidth]{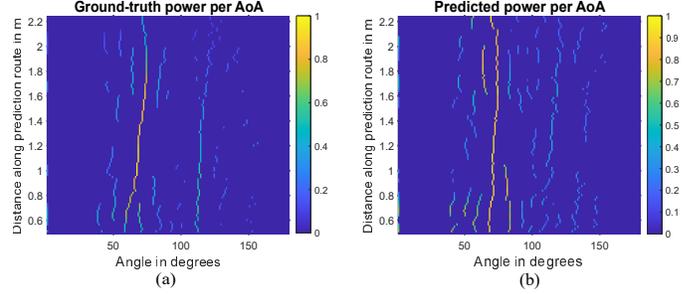}
    \vspace{-0.25in}
    \caption{(a) Ground-truth normalized power per angle for route \#1 marked in Fig.~\ref{fig_chem_area}. The plot shows the power per angle (color-coded) for the rays arriving at the location across the route that is at the distance indicated by the corresponding value on the y-axis. (b) Predicted normalized power per angle on the prediction route. See the color pdf for optimal viewing.}
    \vspace{-0.1in}
    \label{fig_chem_area_aoa1}
\end{figure}

In this section, we discuss our prediction performance in more complex areas that consist of many objects/structures. Consider Area \#2 shown in Fig.~\ref{fig_chem_area}. There are several reflecting objects in this area, including two buildings, pillars, walls/doors, a trash bin, and many other miscellaneous things. The signal rays from the Tx interact with these objects and propagate through the area. The prediction area is of dimensions 8~m~$\times$~3.5~m, \rev{and is} marked on the figure. We first consider one sample prediction route in this area, marked as route \#1 in the figure. In Fig.~\ref{fig_chem_area_result2}, we show the comparison between the ground-truth received signal power and the predicted signal power using our proposed approach, on prediction route \#1. As can be seen, our approach can accurately predict the channel variations, including the small-scale variations due to multipath in the area. Next, in order to evaluate the performance of our ray makeup prediction framework, we predict the detailed power-per-angle profile of the received signal along the prediction route and compare it with the ground truth. More specifically, Fig.~\ref{fig_chem_area_aoa1}~(a) shows the ground-truth normalized power profile across various angles, while Fig.~\ref{fig_chem_area_aoa1}~(b) shows the corresponding angular profile estimated using the predicted signal power on that route. As can be seen, our approach can accurately predict all the rays and their detailed makeup along the route.

\begin{figure}
    \centering
    \includegraphics[width=0.95\linewidth]{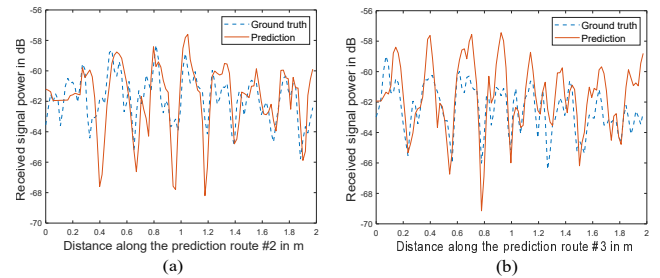}
    \vspace{-0.1in}
    \caption{Channel prediction results, across (a) route \#2 and \rev{(b)} route \#3 from Area \#2 of Fig.~\ref{fig_chem_area}.}
    \label{fig_chem_area_multiple_results}
    \vspace{-0.15in}
\end{figure}

We further evaluate our approach on two additional routes (routes \#2 and \#3) within prediction Area \#2 of Fig.~\ref{fig_chem_area}. As can be seen from Fig.~\ref{fig_chem_area_multiple_results}, our approach can predict the details well for these routes as well.

\begin{figure}
    \centering
    \includegraphics[width=1\linewidth]{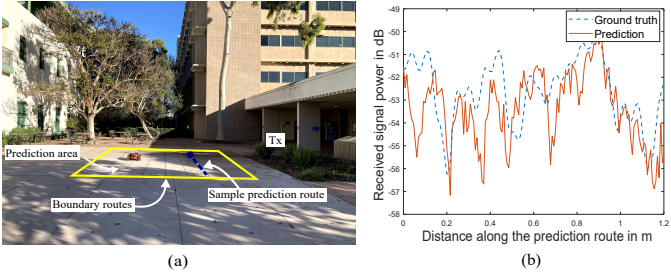}
    \vspace{-0.1in}
    \caption{(a) Area \#3 of dimensions 4.26~m~$\times$~4.26~m -- prior boundary routes and a sample prediction route \rev{are} marked on the figure. (b) Channel prediction across a sample route marked in (a), \rev{starting at the filled circle on the route}.}
    \vspace{-0.1in}
    \label{fig_lawn_area}
\end{figure}

\begin{figure}
    \centering
    \includegraphics[width=1\linewidth]{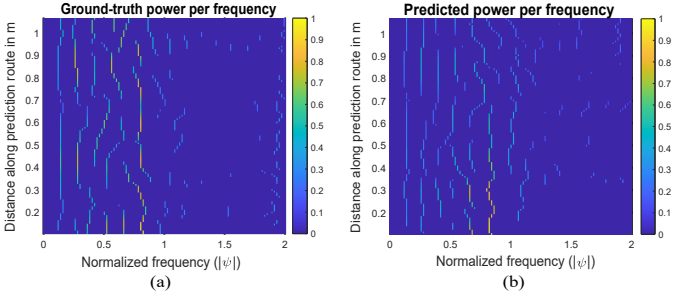}
    \vspace{-0.2in}
    \caption{(a) Ground-truth normalized power spectrum across the sample route marked in Fig.~\ref{fig_lawn_area}~(a). The x-axis shows the normalized frequency \rev{content} ($|\psi|$) at the corresponding distance across the route noted along the y-axis. The colormap shows the normalized power of the peak at the corresponding frequency in the spectrum.  (b) Predicted normalized power per frequency on the prediction route. See the color pdf for optimal viewing.}
    \vspace{-0.15in}
    \label{fig_lawn_area_spectrum}
\end{figure}

We next validate our proposed approach in a more complex area (Area \#3), as shown in Fig.~\ref{fig_lawn_area}~(a). 
This area has many different objects all around that can interact with the signal rays, making the prediction more challenging than the previous two areas. 
The robot then collects the prior power measurements along the marked boundary routes. The prediction area is 4.26~m~$\times$~4.26~m, as marked in the figure. Fig.~\ref{fig_lawn_area}~(b) shows the channel prediction result along a sample route, using our proposed framework. As can be seen, the prediction matches the detailed variations across the route well. 

In order to further evaluate the performance of our ray makeup prediction framework, we predict the power-per-frequency profile (or the power spectrum) of the received signal along the sample prediction route. Fig.~\ref{fig_lawn_area_spectrum}~(a) shows the ground truth power spectrum as a function of the normalized frequency $|\psi| = |\cos \phi_{\text{Tx}} - \cos \phi|$ along the route, where $\phi$ is the AoA of the incoming ray. Fig.~\ref{fig_lawn_area_spectrum}~(b) then shows the predicted power spectrum, which matches the ground truth well, thus validating our approach for predicting the detailed ray makeup at unvisited locations. Note that in Fig.~\ref{fig_lawn_area_spectrum}, we showed the power profile as a function of $|\psi|$, and not \rev{as a function of} the angle $\phi$. This is due to the fact that Area \#3 is a very complex area with objects located all around the prediction region, making it challenging to estimate the ground-truth angles from just the ground-truth signal power measurements on the prediction route. Thus, in order to validate our prediction framework and get around the lack of knowledge of ground-truth angles, we instead show the power profile as a function of the normalized frequency $|\psi|$, which is a direct function of the AoAs. \rev{For Area \#2 of Fig.~\ref{fig_chem_area}, on the other hand, we showed the power per angle profile directly as a function of the angles, since the objects were on one side of the area and we could thus deduce the ground-truth angles.}

Overall, our \rev{experimental} results confirm that our proposed approach can accurately predict the wireless channel power as well as the detailed makeup of the rays at unvisited locations in an area, without the need for knowing the material property/geometry of the objects in the area.

\begin{figure}
    \centering
    \includegraphics[width=1\linewidth]{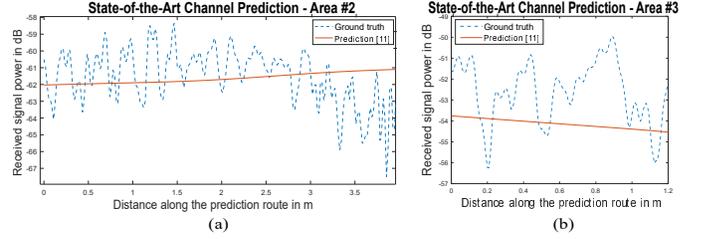}
    \vspace{-0.2in}
    \caption{State-of-the-art prediction \rev{framework of} \cite{malmirchegini2012spatial} on (a) prediction route \#1 of Area \#2 of  Fig.~\ref{fig_chem_area} and (b) prediction route of Area \#3 of Fig.~\ref{fig_lawn_area}~(a).} 
    \vspace{-0.15in}
    \label{fig_sota_chem}
\end{figure}

\vspace{-0.1in}
\subsection{Comparison with State-of-the-Art Prediction Framework}
We next implement the prediction framework \rev{of \cite{malmirchegini2012spatial}} on our experimental data and compare its performance with our proposed approach. \fin{We chose \cite{malmirchegini2012spatial} to compare with as its approach is heavily utilized in this area \cite{fink2013robust, caccamo2017rcamp, kalogerias2018spatially, ghaffarkhah2014dynamic, muralidharan2021communication}. Moreover, as discussed in Sec.~\ref{sec_intro}, other prediction work either need object information (or make assumptions about properties of the objects), or use machine learning, albeit in a simulation environment}. \rev{As discussed in Sec.~\ref{sec_intro}, \cite{malmirchegini2012spatial} uses probabilistic methods to predict the shadowing/path loss components of the channel.}

Consider prediction on route \#1 of Fig.~\ref{fig_chem_area}. Fig.~\ref{fig_sota_chem}~(a) shows the performance of this state-of-the-art approach on this route. \rev{It} can be seen that \rev{while} the signal mean is predicted well, detailed variations arising from multipath \rev{are not captured}. \rev{This is as expected since \cite{malmirchegini2012spatial} predicts the path loss and shadowing components of the signal but not the multipath variations}. \fin{As such, it is more suitable for areas dominated by path loss and shadowing.} 
\rev{Next}, Fig.~\ref{fig_sota_chem}~(b) shows the performance of \cite{malmirchegini2012spatial} on the sample prediction route marked in Fig.~\ref{fig_lawn_area}~(a). Similarly, the detailed multipath variations cannot be captured. Our approach, on the other hand, is fundamentally different, as it focuses on the rays passing through the area and shows how to predict the detailed makeup of the rays.

Overall, our proposed framework and results show its possible to predict detailed ray makeup and the subsequent detailed channel variations at unvisited locations in the workspace. 

\vspace{-0.1in}
\section{Conclusion}\label{sec_conclusions}
In this paper, we have considered the problem of robotic wireless channel prediction, and proposed a new framework for predicting the detailed ray makeup (as well as the resulting received channel power) at any unvisited location in an area of interest, using only a small number of prior received power measurements collected by an unmanned vehicle. More specifically, we have shown how to methodically design the prior robot route, via an enclosure-based approach, such that it can capture the key information content of the rays propagating through the area. We then showed how we can use the measurements along the designed boundary routes and fully predict the detailed makeup of the rays at all other unvisited locations in the area of interest. It is noteworthy that our approach does not need any knowledge on the geometry or the material property of the objects in the area. Finally, we validated our proposed approach through extensive experiments in three different areas and showed that it can accurately predict the key parameters of the rays, as well as the detailed signal power variations.  

\appendices
\vspace{-0.1in}
\section{}\label{app_ground_path}
As can be seen from Eq.~\ref{eq_point_signal}, the term $\alpha_{\text{Tx}}^2$ is monotonically decreasing with increasing distance $l_{\text{Tx}}$.
Next, consider the ground path terms in Eq.~\ref{eq_channel_power} at an antenna array:
\begin{align}\label{eq_ground_terms}
    |c_g(d)|^2 &=  \alpha_g^2 + \alpha_{\text{Tx}} \alpha_g \left\{ e^{j\frac{2\pi}{\lambda} [l_{\text{Tx}} - d\cos \phi_{\text{Tx}} -l_g(d)]} \right. \nonumber \\ 
    & \left. + e^{-j\frac{2\pi}{\lambda} [l_{\text{Tx}} - d\cos \phi_{\text{Tx}} -l_g(d)]} \right\}.
\end{align}
The first term in the equation, $\alpha_g^2$, is a smoothly varying term \fin{in far-field scenarios}. Furthermore, it is weak when compared to $\alpha_{\text{Tx}}^2$, and varies much slower than the variations that we observe in the signal mean. Thus it is not a major contributor to the variations in the mean. Next, consider the second term, whose frequency content depends on the difference in the path lengths of the direct and ground paths. The amplitude of this term is also significant since it contains the impact of $\alpha_{\text{Tx}}$, which is a dominant term. Rewriting the phase expression in this term, $l_g(d)$ can be written as a function of the distance $d$ along the array as $l_g(d) = l_g(0) - d \cos \theta_{\text{arr}}$. The frequency content of the second term in Eq.~\ref{eq_ground_terms}, normalized with respect to wavelength, can be written as a function of the AoAs of the paths as $\psi_g = \cos \phi_\text{Tx} - \cos \theta_{\text{arr}}$, where $\theta_\text{arr}$ is the angle-of-arrival of the ground path with respect to the antenna array. 

Thus, it is straight-forward to calculate these two angles, for any positions of the Tx and Rx. 
Through geometry, we can see that the difference in angles increases monotonically, as $\phi_{\text{Tx}}$ is varied from $90^{\circ}$ to $0^{\circ}$. Furthermore, while it is not obvious from the expression for $\psi_g$, it can indeed be verified through geometry that $\psi_g$ increases monotonically from $0$ to $\left[1 - \cos \left( \tan^{-1} (2 h_a / l_\text{Tx}) \right)\right]$, over the same range of angles for $\phi_{\text{Tx}}$. Thus, $\left[1 - \cos \left( \tan^{-1} (2 h_a / l_\text{Tx}) \right)\right]$ is the maximum possible frequency content of the second term in Eq.~\ref{eq_ground_terms}. One can easily extend this analysis for $\phi_{\text{Tx}} \in [90^\circ, 180^\circ]$. 

To get a sense of the range of $|\psi_g|$, consider a typical scenario with $h_a = 0.5$~m, and $l_\text{Tx} = 5$~m. $|\psi_g|$ then has a maximum value of $0.02$, which only reduces further with increasing $l_\text{Tx}$, as is observed through simulations as well. In comparison, $|\psi|$ values for other signal paths can have a maximum value of $2$, which is considered a high-frequency variation in the signal. Hence, the low frequency variations observed in the signal across a route are a result of the ground path interference with the direct path from the Tx.

\vspace{-0.1in}
\bibliographystyle{IEEEtran}
\bibliography{ref}

\end{document}